\documentclass[amsmath,floatfix,twocolumn,superscriptaddress]{revtex4-1}
\usepackage{subfigure}
\usepackage{siunitx}
\usepackage{amssymb}
\usepackage{amsmath}
\usepackage{graphicx}
\usepackage{array}
\usepackage{dcolumn}
\usepackage{psfrag}
\usepackage{bm}
\usepackage{color}
\usepackage{multirow}

\begin{document}

\title{Giant Electron-Phonon Coupling Induced Band-Gap Renormalization in Anharmonic Silver Chalcohalide Antiperovskites}

\author{Pol Benítez}
    \affiliation{Group of Characterization of Materials, Departament de F\'{i}sica, Universitat Polit\`{e}cnica de Catalunya, Campus Diagonal Bes\`{o}s, Av. Eduard Maristany 10--14, 08019 Barcelona, Spain}
    \affiliation{Research Center in Multiscale Science and Engineering, Universitat Politècnica de Catalunya, Campus Diagonal-Bes\`{o}s, Av. Eduard Maristany 10--14, 08019 Barcelona, Spain}

\author{Siyu Chen}
    \affiliation{Department of Materials Science and Metallurgy, University of Cambridge, Cambridge CB30FS, UK}
    \affiliation{Cavendish Laboratory, University of Cambridge, Cambridge CB30HE, UK}

\author{Ruoshi Jiang}
    \affiliation{Department of Materials Science and Metallurgy, University of Cambridge, Cambridge CB30FS, UK}

\author{Cibrán López}
    \affiliation{Group of Characterization of Materials, Departament de F\'{i}sica, Universitat Polit\`{e}cnica de Catalunya, Campus Diagonal Bes\`{o}s, Av. Eduard Maristany 10--14, 08019 Barcelona, Spain}
    \affiliation{Research Center in Multiscale Science and Engineering, Universitat Politècnica de Catalunya, Campus Diagonal-Bes\`{o}s, Av. Eduard Maristany 10--14, 08019 Barcelona, Spain}

\author{Josep-Llu\'is Tamarit}
    \affiliation{Group of Characterization of Materials, Departament de F\'{i}sica, Universitat Polit\`{e}cnica de Catalunya, Campus Diagonal Bes\`{o}s, Av. Eduard Maristany 10--14, 08019 Barcelona, Spain}
    \affiliation{Research Center in Multiscale Science and Engineering, Universitat Politècnica de Catalunya, Campus Diagonal-Bes\`{o}s, Av. Eduard Maristany 10--14, 08019 Barcelona, Spain}

\author{Jorge Íñiguez-González}
    \affiliation{Materials Research and Technology Department, Luxembourg Institute of Science and Technology (LIST), Avenue des Hauts-Fourneaux 5, L-4362 Esch/Alzette, Luxembourg}
    \affiliation{Department of Physics and Materials Science, University of Luxembourg, 41 Rue du Brill, L-4422 Belvaux, Luxembourg}

\author{Edgardo Saucedo}
	 \affiliation{Research Center in Multiscale Science and Engineering, Universitat Politècnica de Catalunya, Campus Diagonal-Bes\`{o}s, Av. Eduard Maristany 10--14, 08019 Barcelona, Spain}
	\affiliation{Micro and Nanotechnologies Group, Emerging Thin Film Photovoltaics Lab, Departament d’Enginyeria Electr\`{o}nica, Universitat 
	Polit\`{e}cnica de Catalunya, Campus Diagonal Bes\`{o}s, Av. Eduard Maristany 10-14, 08019 Barcelona, Spain.}

\author{Bartomeu Monserrat}
    \affiliation{Department of Materials Science and Metallurgy, University of Cambridge, Cambridge CB30FS, UK}
    \affiliation{Cavendish Laboratory, University of Cambridge, Cambridge CB30HE, UK}

\author{Claudio Cazorla}
    \affiliation{Group of Characterization of Materials, Departament de F\'{i}sica, Universitat Polit\`{e}cnica de Catalunya, Campus Diagonal Bes\`{o}s, Av. Eduard Maristany 10--14, 08019 Barcelona, Spain}
    \affiliation{Research Center in Multiscale Science and Engineering, Universitat Politècnica de Catalunya, Campus Diagonal-Bes\`{o}s, Av. Eduard Maristany 10--14, 08019 Barcelona, Spain}

\begin{abstract}

{\bf Abstract.} Silver chalcohalide antiperovskites (CAP), Ag$_{3}$XY (X = S, Se; Y = Br, I), are a family of highly anharmonic 
inorganic compounds with great potential for energy applications. However, a substantial and unresolved discrepancy exists 
between the optoelectronic properties predicted by theoretical first-principles methods and those measured experimentally 
at room temperature, hindering the fundamental understanding and rational engineering of CAP. In this work, we employ 
density functional theory, tight-binding calculations, and anharmonic Fr\"ohlich theory to investigate the optoelectronic 
properties of CAP at finite temperatures. Near room temperature, we observe a giant band-gap ($E_{g}$) reduction of approximately 
$20$--$60$\% relative to the value calculated at $T = 0$~K, bringing the estimated $E_{g}$ into excellent agreement with 
experimental measurements. This relative $T$-induced band-gap renormalization is roughly twice the largest value previously 
reported in the literature for similar temperature ranges. Low-energy optical polar phonon modes, which break inversion symmetry 
and promote the overlap between silver and chalcogen $s$ electronic orbitals in the conduction band, are identified as the 
primary contributors to this giant $E_{g}$ reduction. Furthermore, when considering temperature effects, the optical absorption 
coefficient of CAP increases by nearly an order of magnitude for visible light frequencies. These insights not only bridge 
a crucial gap between theory and experiment but also open pathways for future technologies where temperature, electric fields, 
or light dynamically tailor optoelectronic behavior, positioning CAP as a versatile platform for next-generation energy 
applications.\\


\end{abstract}

\maketitle

\section*{Introduction}
\label{sec:intro}
Electron-phonon coupling (EPC), arising from the interactions between electrons and lattice vibrations, is ubiquitous 
in materials and responsible for a wide range of condensed matter physical effects \cite{epc1,epc2,epc3}. For example, 
EPC plays a crucial role in the temperature ($T$) dependence of electrical resistivity in metals, carrier mobility in 
semiconductors, optical absorption in indirect band gap semiconductors, and the onset of conventional superconductivity. 
Additionally, EPC enables the thermalization of hot carriers, influences the phonon dispersion in metals, and determines 
the $T$-dependence of electronic energy bands in solids \cite{monserrat18,monserrat17}.

Likewise, the band gap ($E_{g}$) of semiconducting and dielectric materials can be significantly affected by EPC, typically 
decreasing with increasing temperature (the so-called Varshni effect \cite{varshni67}). This common $E_{g}$ behavior can 
be explained by the Allen-Heine-Cardona perturbative theory, which attributes it to a larger $T$-induced energy increase 
in the valence band compared to the conduction band due to a greater sensitivity to phonon population variations (i.e., 
larger second-order electron-phonon coupling constants) \cite{ahc1,ahc2,ahc3}. Representative examples of this thermal 
$E_{g}$ dependence include diamond, which exhibits a $\sim 5$\%~ band-gap reduction at around $1000$~K \cite{diamond1,diamond2}; 
antimony sulfide (Sb$_{2}$S$_{3}$), which shows a $E_{g}$ reduction of $200$~meV in the temperature range $10 \le T \le 
300$~K \cite{antimony}; and molecular crystals, which display record band-gap reductions of $15$--$20$\%~ at low 
temperatures \cite{molecular}. To note that \textit{anomalous} band-gap thermal behaviour, in which $E_{g}$ increases with 
increasing temperature, has also been observed in a variety of materials such as black phosphorus \cite{anomal1}, halide 
perovskites \cite{anomal2}, and chalcopyrite \cite{anomal3} and hydride \cite{diamond2} compounds.

\begin{figure*}[t]
    \centering
    \includegraphics[width=1.0\linewidth]{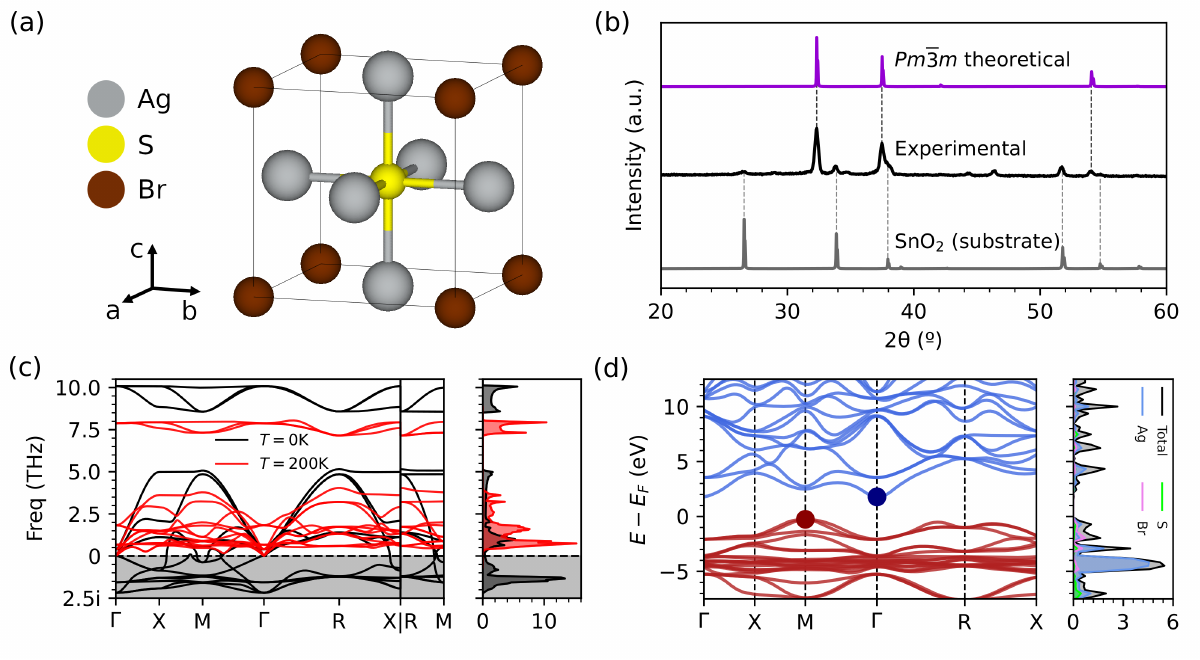}
	\caption{\textbf{General physical properties of the archetypal CAP Ag$_{3}$SBr.}
	(a)~The cubic $Pm\overline{3}m$ phase experimentally observed at room temperature. 
	(b)~Experimental diffractogram of Ag$_3$SBr \cite{cano2024} compared with the theoretical one estimated for 
	the cubic $Pm\overline{3}m$ phase. 
	(c)~Vibrational phonon spectrum (left) and phonon density of states (right) calculated within the harmonic 
	approximation for the cubic $Pm\overline{3}m$ phase of Ag$_3$SBr at $T = 0$~K (black lines) and at $T = 200$~K 
	(red lines) fully considering anharmonic effects. (d)~Electronic band structure (left) and density of states (right) 
	of Ag$_3$SBr calculated with the hybrid functional HSEsol \cite{hse06}. Red and blue lines (dots) represent valence 
	(top of the valence) and conduction (bottom of the conduction) bands, respectively.
	}
    \label{fig1}
\end{figure*}

Highly anharmonic silver chalcohalide antiperovskites (CAP) \cite{benitez24} with chemical formula Ag$_{3}$XY (X = S, Se;
Y = Br, I) are structurally similar to the lead halide perovskites (e.g., CsPbI$_{3}$), with the ``anti'' designation
indicating the exchange of anions and cations compared to the typical ionic perovskite arrangement. Analogous to lead
halide perovksites, CAP are highly promising materials for energy and optoelectronic applications \cite{takahashi66,hull04,
wakamura90,sakuma85,kawamura80,magistris72}, offering low toxicity due to their lead-free composition \cite{palazon22,
ghorpade23}. The two most extensively studied CAP compounds, Ag$_{3}$SBr and Ag$_{3}$SI, possess experimentally determined
band gaps of approximately $1.0$~eV \cite{luna23,cano2024}, making them favorable for photovoltaic applications.
These materials have also been recognized as high temperature superionic conductors \cite{takahashi66,hull04}. Additionally, 
CAP have been investigated as potential thermoelectric materials \cite{kawamura80,magistris72} owing to their substantial 
vibrational anharmonicity and unique charge transport properties \cite{wakamura90,sakuma85}.

Intriguingly, for both Ag$_{3}$SBr and Ag$_{3}$SI, there is a enormous disagreement between the $E_{g}$ predicted by
first-principles methods (at $T = 0$~K under static lattice conditions) and those measured experimentally at room 
temperature. In particular, high-level density functional theory (DFT) calculations employing hybrid functionals and 
including spin-orbit coupling (SOC) effects estimate the band gap of these two archetypal CAP to be $1.8$ and $1.4$~eV, 
respectively \cite{luna23,cano2024,liu21}. The $E_{g}$ discrepancies between theory and measurements amount to $60$--$80$\%~ 
of the experimental values (i.e., differences of $0.5$--$0.8$~eV), which are unusually large and call for a careful 
inspection of the factors causing them.

In this study, we assessed the influence of EPC and temperature effects on the $E_{g}$ and optical absorption spectra of CAP 
using first-principles DFT methods, tight-binding calculations, and anharmonic Fr\"ohlich theory. Near room temperature, our 
computational investigations revealed a giant $E_{g}$ reduction of $20$--$60$\% relative to the value calculated at $T = 0$~K, 
bringing the estimated band gap into excellent agreement with the experimental values. Low-energy optical polar phonons, which 
cause large symmetry-breaking structural distortions and promote the overlap between silver and chalcogen $s$ electronic 
orbitals in the conduction band, were identified as the primary mechanism driving this substantial $T$-induced band-gap 
reduction. Furthermore, at finite temperatures the optical absorption spectra of CAP were significantly enhanced, in some cases 
by nearly an order of magnitude. The polar nature of the phonons causing these effects opens up new technological possibilities, 
where the optoelectronic properties of materials could be effectively manipulated by external electric fields and light.

\begin{figure*}[t]
    \centering
    \includegraphics[width=1.0\linewidth]{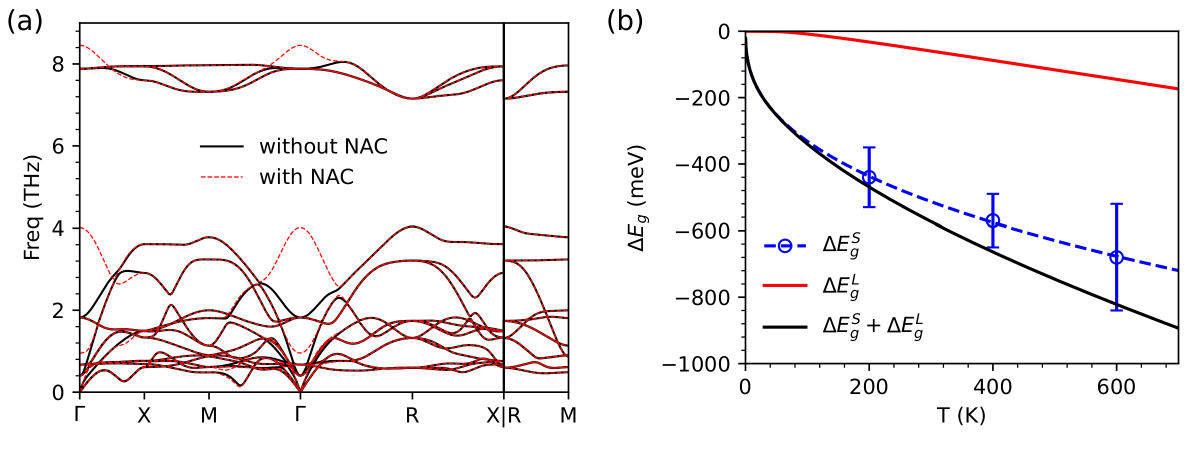}
	\caption{\textbf{Anharmonic phonon spectrum and thermal band-gap corrections estimated for the archetypal CAP 
	Ag$_{3}$SBr.} (a)~Anharmonic phonon spectrum obtained at $T = 200$~K neglecting (black solid lines) and considering 
	(red dashed lines) non-analytical corrections (NAC). (b)~Short- and long-wavelength phonon band-gap corrections, 
	$\Delta E_{g}^{S}$ and $\Delta E_{g}^{L}$, respectively, expressed as a function of temperature (excluding quantum 
	nuclear effects). The short-range correction term was evaluated at several temperature points (blue circles and 
	error bars); a power law function was subsequently fitted to the $\Delta E_{g}^{S}$ data (blue dashed line) as a 
	guide to the eye. Calculations were performed at the HSEsol+SOC level \cite{hse06}.}
    \label{fig2}
\end{figure*}

\section*{Results}
\label{sec:results}
The room-temperature phase of both Ag$_{3}$SI and Ag$_{3}$SBr have been experimentally identified as cubic with space 
group $Pm\overline{3}m$ \cite{cano2024,sakuma80,hoshino81,cho94}. This phase is characterized by a five-atoms unit cell: 
a chalcogen atom in the center of the cube, halide atoms in each vertex, and silver atoms in the center of each face 
(Figs.~\ref{fig1}a-b). Phonon calculations of this phase within the harmonic approximation ($T = 0$~K) reveal imaginary 
phonon branches, thus indicating dynamical instability. However, when phonons are calculated fully accounting for anharmonic 
effects at finite-$T$ conditions, the resulting phonon spectrum is well-behaved with no signs of instabilities (Fig.~\ref{fig1}c) 
\cite{benitez24}. Thus, the cubic $Pm\overline{3}m$ phase was considered throughout this work for all CAP.

As discussed in the Introduction, the discrepancies between the experimentally measured (at $T = 300$~K) and theoretically 
determined (at $T = 0$~K) band gaps of Ag$_{3}$SBr and Ag$_{3}$SI are tremendously large (i.e., $60$--$80$\% of the experimental 
values). Therefore, we investigated their potential causes by assessing the impact of electron-phonon coupling (EPC) and 
temperature on the band gap of CAP. To this end, we performed first-principles calculations and \emph{ab initio} molecular 
dynamics (AIMD) simulations based on DFT (Methods). Additionally, to capture long-range EPC effects, we employed anharmonic 
Fr\"ohlich theory \cite{frohlich1,frohlich2,frohlich3,frohlich4} considering long-range dipole-dipole interactions and 
$T$-renormalized phonons (Methods). Furthermore, the optical absorption spectra of all CAP were assessed at $T \neq 0$~K 
conditions and the main EPC mechanisms underlying the $E_{g}$ discrepancies were identified with the help of a tight-binding 
model.   
\\

\begin{table*}[t]
    \centering
    \begin{tabular}{cccccccccccc}
    \hline
    \hline
	    & & & & & & & & & & & \\
	    CAP & $E_{g}^{0{\rm K}}$ \qquad & \qquad $ E_{g}^{200{\rm K}} $ & $\Delta E_{g}^{S}$ & $\Delta E_{g}^{L}$ \qquad & \qquad $ E_{g}^{400{\rm K}} $ & $\Delta E_{g}^{S} $ & $\Delta E_{g}^{L} $ \qquad & \qquad $ E_{g}^{600{\rm K}} $ & $\Delta E_{g}^{S} $ & $\Delta E_{g}^{L} $ \qquad & \qquad $ E_{g}^{\rm exp}$ \\
	    & [eV] \qquad & \qquad [eV] & [meV] & [meV] \qquad & \qquad [eV] & [meV] & [meV] \qquad & \qquad [eV] & [meV] & [meV] \qquad & \qquad [eV] \\
	    & & & & & & & & & & & \\
    \hline
	    & & & & & & & & & & & \\
	    Ag$_3$SBr  & 1.8 \qquad & \qquad 1.3$\pm$0.1 & $ -440 $ & $ -32 $ \qquad & \qquad 1.1$\pm$0.1 & $ -570 $ & $ -88 $ \qquad & \qquad 1.0$\pm$0.2 & $ -680 $ & $ -145 $ \qquad & \qquad 1.0 \\
            Ag$_3$SI   & 1.4 \qquad & \qquad 1.1$\pm$0.1 & $ -260 $ & $ -21 $ \qquad & \qquad 1.1$\pm$0.1 & $ -230 $ & $ -58 $ \qquad & \qquad 0.8$\pm$0.1 & $ -490 $ & $ -95 $ \qquad  & \qquad 0.9 \\
            Ag$_3$SeBr & 1.6 \qquad & \qquad 0.9$\pm$0.1 & $ -630 $ & $ -32 $ \qquad & \qquad 0.7$\pm$0.1 & $ -770 $ & $ -82 $ \qquad & \qquad liquid    \qquad  & \qquad $ - $ & $ - $ \qquad & \qquad $ - $  \\
            Ag$_3$SeI  & 1.3 \qquad & \qquad 0.9$\pm$0.1 & $ -370 $ & $ -24 $ \qquad & \qquad 0.8$\pm$0.2 & $ -400 $ & $ -61 $ \qquad & \qquad liquid    \qquad  & \qquad $ - $ & $ - $ \qquad & \qquad $ - $  \\
	    & & & & & & & & & & & \\
    \hline
    \hline
    \end{tabular}
	\caption{{\bf Theoretical band gaps of CAP as a function of temperature.} $E_{g}$ values were obtained at zero 
	temperature (excluding quantum nuclear effects) at $T = 200, 400$, and $600$~K. Calculations were performed 
	at the HSEsol+SOC level \cite{hse06}. Numerical uncertainties are provided, which mainly result from the 
	$\Delta E_{g}^{S}$ correction term. Short- and long-wavelength phonon band-gap corrections, $\Delta E_{g}^{S}$ 
	and $\Delta E_{g}^{L}$, respectively, are provided at each temperature (excluding quantum nuclear effects). The 
	experimental band gaps measured at room temperature for Ag$_3$SBr and Ag$_3$SI \cite{cano2024} are shown for comparison.}
    \label{tab:bandgap}
\end{table*}

\begin{figure*}[t]
    \centering
    \includegraphics[width=1.0\linewidth]{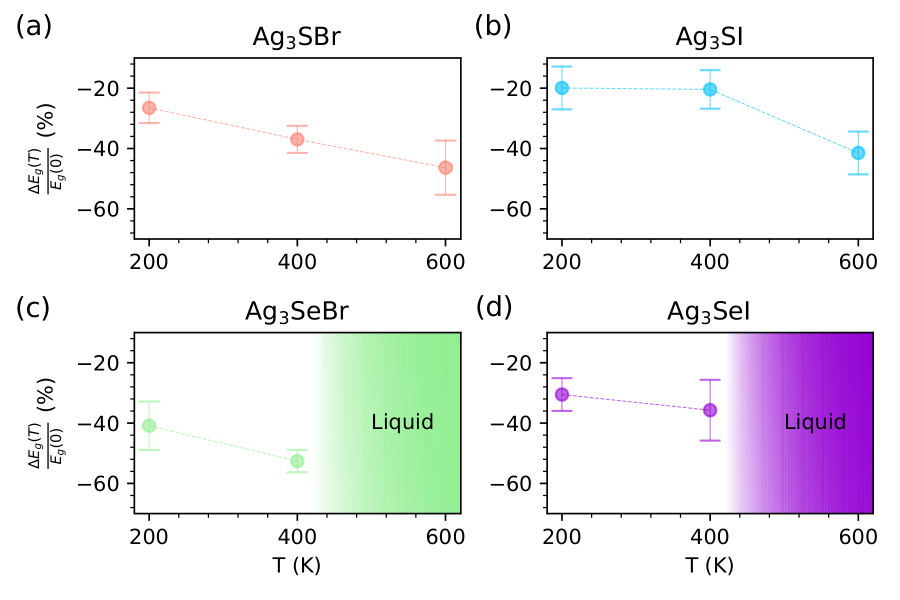}
        \caption{\textbf{Temperature-induced relative band-gap variation in CAP.}
                Percentages are referenced to the band gap calculated at $T = 0$~K conditions (excluding quantum nuclear
                effects), namely, $\Delta E_{g}(T) = E_{g}(T) - E_{g}(0)$, for (a)~Ag$_3$SBr, (b)~Ag$_3$SI, (c)~Ag$_3$SeBr,
                and (d)~Ag$_3$SeI. Error bars indicate numerical uncertainties and dashed lines are guides to the eye.
		Shaded areas indicate regions of thermodynamic stability of the liquid phase (theory). Calculations were 
		performed at the HSEsol+SOC level \cite{hse06}.}
    \label{fig3}
\end{figure*}

\begin{figure*}[t]
    \centering
    \includegraphics[width=1.0\textwidth]{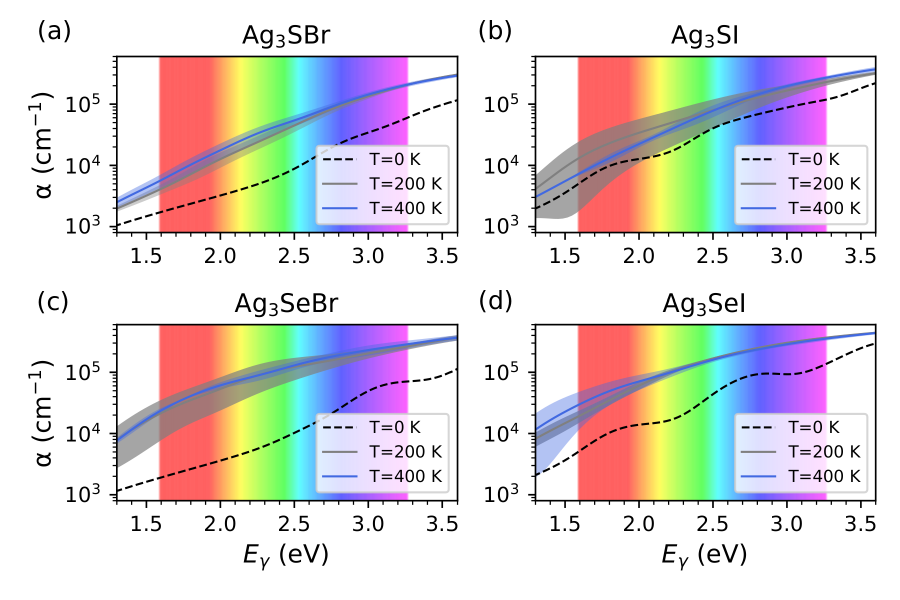}
	\caption{\textbf{Optical absorption coefficient ($\alpha$) of CAP calculated at different temperatures as a function 
	of photon energy.} (a)~Ag$_{3}$SBr, (b)~Ag$_{3}$SI, (c)~Ag$_{3}$SeBr, and (d)~Ag$_{3}$SeI. Solid lines represent the 
	estimated average values and statistical errors are indicated with shaded thick curves. The rainbow-colored region 
	denotes photons with energy in the visible spectrum. Calculations were performed at the HSEsol+SOC level \cite{hse06}.}
    \label{fig4}
\end{figure*}

{\bf EPC band-gap renormalization in CAP.}~The $T$-renormalized band gap of CAP was calculated like:
\begin{equation}
	E_{g} (T) = E_{g} (0) + \Delta E_{g} (T),
\label{eq:eg}
\end{equation}
where $E_{g} (0)$ represents the static band gap and the correction term $\Delta E_{g}$ can be expressed as the sum of 
short- ($S$) and long-wavelength ($L$) phonon contributions \cite{frohlich2}:
\begin{equation}
	\Delta E_{g} (T) = \Delta E_{g}^{S} (T) + \Delta E_{g}^{L} (T).
\label{eq:corr}
\end{equation}

The short-wavelength phonon correction was estimated by performing AIMD simulations with a supercell (Methods) and 
averaging the band-gap value over several generated configurations, namely \cite{frohlich1}:
\begin{equation}
 	\Delta E_{g}^{S} (T) = \frac{1}{N} \sum_{k=1}^{N} E_{g} (\lbrace {\bf R}_{k} (T) \rbrace) - E_{g} (0),
\label{eq:corr-s}
\end{equation}
where $N$ represents the total number of considered configurations and $\lbrace {\bf R}_{k} \rbrace$ the atomic 
positions of the $k$-th configuration. In polar materials, there is an additional contribution to the band-gap 
renormalization stemming from long-range Fr\"ohlich coupling that is not fully captured by the finite size of 
the supercells employed in the AIMD simulations \cite{frohlich1,frohlich2,frohlich3,frohlich4,melo23}. This 
long-wavelength phonon band-gap correction can be expressed as:
\begin{equation}
	\Delta E_{g}^{L} (T) = \Delta \epsilon_{\rm CB}^{L} (T) - \Delta \epsilon_{\rm VB}^{L} (T),  
\label{eq:corr-l}
\end{equation}
where CB and VB refer to the bottom conduction and top valence band levels, respectively. 

For a 3D polar material, the $T$-induced energy level shifts appearing in Eq.~(\ref{eq:corr-l}) can be computed as 
\cite{frohlich1}:
\begin{equation}
	\Delta \epsilon_{\rm i}^{L} (T) = \frac{2 \alpha_{P}}{\pi} \hbar \omega_{\rm LO} \tan^{-1} \left( 
	\frac{q_{\rm F}}{q_{\rm LO,i}} \right) \left[ 2n_{T} + 1 \right],
\label{eq:corr-l-2}
\end{equation}
where $\alpha_{P}$ represents the polaron constant, $\omega_{\rm LO}$ the phonon frequency averaged over the three 
longitudinal optical $\Gamma$ phonon modes \cite{melo23}, and $q_{\rm F}$ a truncation factor. The truncation factor 
$q_{\rm F}$ can be approximated as the Debye sphere radius and $q_{\rm LO,i}$ is defined as $\sqrt{2m^{*} \left( 
\omega_{\rm LO} + \omega_{i} \right) / \hbar}$, $m^{*}$ being the charge carrier effective mass and $\hbar \omega_i$ 
the state energy. The term $n_{T}$ is the Bose-Einstein occupation number corresponding to the average LO 
vibrational frequency, and the polaron constant can be computed as \cite{frohlich1}:
\begin{equation}
\alpha_{P} = \frac{e^{2}}{4\pi \epsilon_{0}\hbar}\left(\frac{1}{\varepsilon_{\infty}}-\frac{1}{\varepsilon_{0}}\right)\left( \frac{m^{*}}{2\hbar \omega_{\text{LO}}} \right)^{1/2},
\label{eq:polaron}
\end{equation}
where $\varepsilon_{\infty}$ is the high-frequency dielectric constant and $\varepsilon_{0}$ the static permittivity of
the system. Quantum nuclear effects have been disregarded throughout this work, hence the $T$-induced energy level shifts 
in Eq.~(\ref{eq:corr-l-2}) were offset by their zero-temperature values $\Delta \epsilon_{\rm i}^{L} (0)$.

Figure~\ref{fig2}a presents the anharmonic phonon spectrum calculated for Ag$_{3}$SBr under finite-temperature conditions, 
accounting for long-range dipole-dipole interactions (i.e., including non-analytical corrections), which result in very
large LO-TO splitting near the reciprocal space point $\Gamma$. Figure~\ref{fig2}b shows the corresponding short- and 
long-wavelength phonon band-gap corrections expressed as a function of temperature, which are always negative. Since in 
this study the $\Delta E_{g}^{L}$ correction term has been calculated using the material's anharmonic phonon spectrum, we 
refer to this method as anharmonic Fr\"ohlich theory (Methods). 

In Fig.~\ref{fig2}b, it is observed that near room temperature the $\Delta E_{g}^{S}$ correction is dominant and significantly 
larger than $\Delta E_{g}^{L}$ , approximately seven times greater in absolute value. Notably, at $T = 400$~K, the total 
band-gap correction for Ag$_{3}$SBr amounts to $0.7$~eV, which is giant, representing roughly $40$\% of the $E_{g}$ value 
calculated at zero temperature (excluding quantum nuclear effects). 

Figure~\ref{fig3} shows the relative band-gap variation, referenced to the value calculated at zero temperature and 
expressed as a function of temperature, for the four CAP compounds Ag$_{3}$SBr, Ag$_{3}$SI, Ag$_{3}$SeBr and Ag$_{3}$SeI. 
In all cases, the band gap significantly decreases as the temperature increases (Table~\ref{tab:bandgap}). The relative 
$T$-induced $E_{g}$ reduction is largest for Ag$_{3}$SeBr and smallest for Ag$_{3}$SI. In particular, near room temperature, 
the band gap of Ag$_{3}$SBr and Ag$_{3}$SI is reduced by $39$\% and $21$\% while those of Ag$_{3}$SeBr and Ag$_{3}$SeI 
decrease by $56$\% and $38$\%, respectively (Fig.~\ref{fig3}). As shown in Table~\ref{tab:bandgap}, the agreement between 
the experimental and theoretical $E_{g}$ values for Ag$_{3}$SBr and Ag$_{3}$SI improves as the temperature increases. In 
Ag$_{3}$SeBr and Ag$_{3}$SeI, the liquid phase is stabilized over the crystal phase at moderate temperatures 
(Fig.~\ref{fig3}c--d), thus no band gaps were estimated for these two compounds at $T > 400$~K conditions. 

Notably, our theoretical $E_{g}$ results obtained at $T = 400$~K are fully consistent with the available experimental data 
obtained at room temperature. This excellent agreement near ambient conditions strongly suggests that the neglect of EPC 
effects is the main reason for the huge theoretical-experimental $E_{g}$ discrepancies discussed in the Introduction. The 
$T$-induced relative band-gap renormalizations found in CAP are giant, ranging from $20$ to $60$\% near room temperature, 
setting a new record previously held by molecular crystals, which exhibited a $15$ to $20$\% band-gap renormalization for 
similar temperature ranges \cite{molecular}.    

\begin{figure*}[t]
    \centering
    \includegraphics[width=1.0\textwidth]{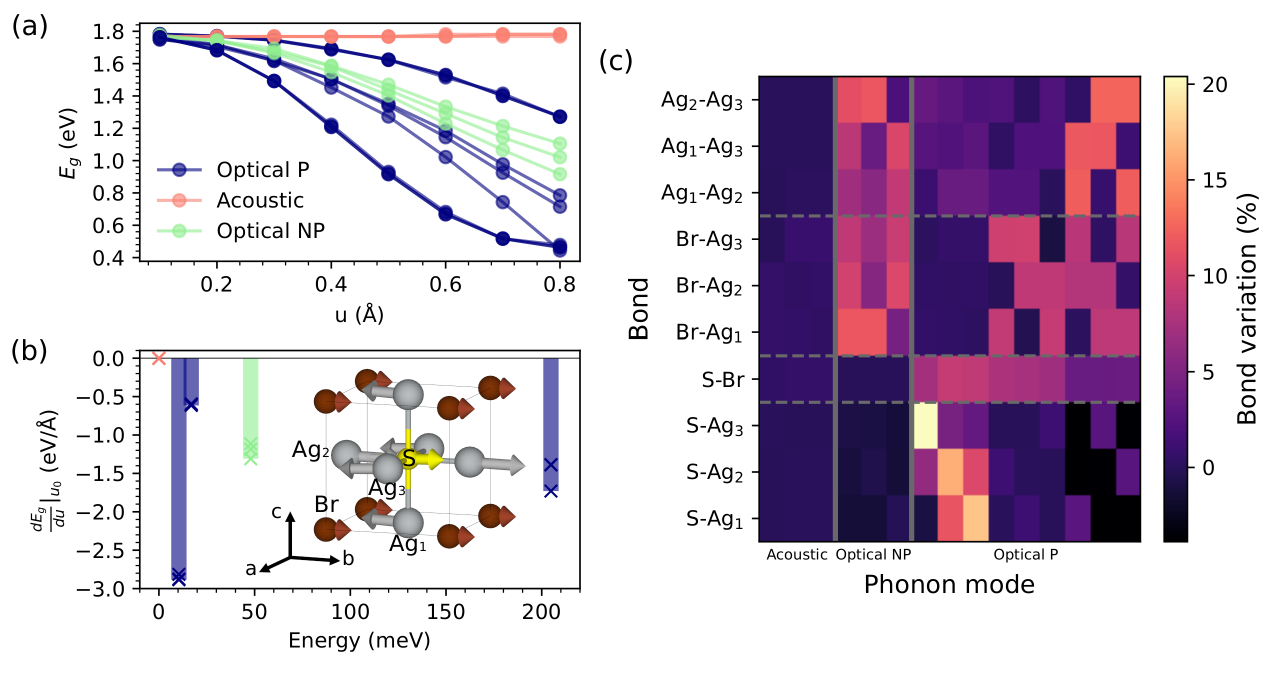}
        \caption{\textbf{Phonon-induced band-gap variation estimated for the archetypal CAP Ag$_{3}$SBr.} (a)~Band gap as a
                function of the lattice distortion amplitude $u$ for acoustic, polar optical (P) and non-polar optical (NP)
                $\Gamma$ phonons. (b)~Derivative of the band gap with respect to the phonon distortion amplitude calculated
                at $u_{0} = 0.4$~\AA~ and expressed as a function of the phonon energy. The eigenmode of the optical polar
                $\Gamma$ phonon rendering the largest band-gap derivative in absolute value is sketched: Ag, S and Br atoms
                are represented with grey, yellow and brown spheres, respectively. Calculations were performed at the 
		HSEsol+SOC level \cite{hse06}. (c)~$\Gamma$ phonon-induced relative bond length distortions in the cubic 
		$Pm\overline{3}m$ phase for a distorsion amplitude of $0.4$~\AA.}
    \label{fig5}
\end{figure*}

Table~\ref{tab:bandgap} also presents the value of the $\Delta E_{g}^{S}$ and $\Delta E_{g}^{L}$ correction terms
estimated for each CAP at three different temperatures. In all cases, both the short- and long-wavelength phonon
corrections are negative, with the former term considerably surpassing the latter in absolute value. For example,
at $T = 400$~K, the short-range band-gap corrections are nine and four times larger than the long-range ones calculated
for Ag$_{3}$SeBr and Ag$_{3}$SI, respectively. As the temperature is raised, the size of the two band-gap correction
terms increase in absolute value, with $|\Delta E_{g}^{L}|$ exhibiting the largest relative enhancement (e.g., 
approximately a $450$\% increase for Ag$_{3}$SBr from $200$ to $600$~K).
\\

\begin{figure*}[t]
    \centering
    \includegraphics[width=1.0\textwidth]{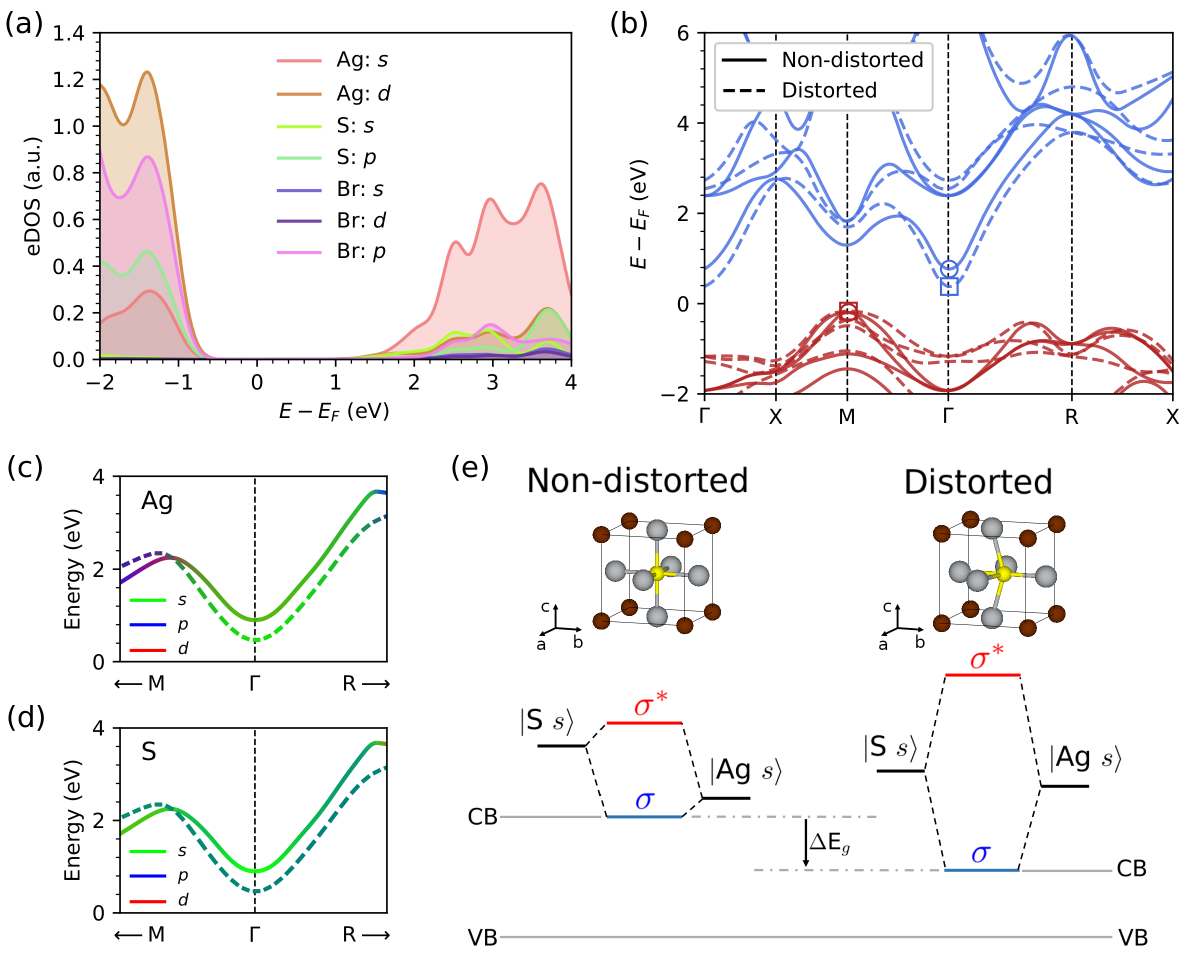}
        \caption{\textbf{Electronic band structure properties of the archetypal CAP Ag$_{3}$SBr.} (a)~Electronic density
        of states calculated for the equilibrium cubic $Pm\overline{3}m$ phase. Partial contributions from different
        atomic orbitals are indicated. Calculations were performed at the HSEsol+SOC level \cite{hse06}. (b)~Electronic 
	band structure calculated for the equilibrium and phonon distorted (i.e., lowest-energy $\Gamma$ optical P mode) 
	cubic $Pm\overline{3}m$ phase; in both cases, the energy bands are referred to a same energy origin, namely, a 
	deep core electronic level that remains unaffected by the distortion (Methods). (c) and (d)~The conduction band 
	near its minimum at $\Gamma$ computed with a TB model for silver and sulfur atoms (Methods); solid and dashed lines 
	correspond to the equilibrium and distorted structures, respectively. (e)~Sketch illustrating the mechanism of 
	band-gap closure induced by low-energy polar soft phonon modes in CAP. Upon phonon distortion, the hybridization 
	of silver and sulfur $s$ electrons in the conduction band is enhanced, lowering (increasing) the energy of the 
	corresponding bonding $\sigma$ (antibonding $\sigma^{*}$) state.} 
        \label{fig6}
\end{figure*}

{\bf $T$-effects on the optical absorption coefficient of CAP.}~Following a similar approach to that used for the calculation 
of $T$-renormalized band gaps (i.e., performing AIMD simulations with a supercell and averaging the quantity of interest 
over several of the generated configurations), we determined the frequency-dependent complex dielectric tensor of CAP under 
$T \neq 0$~K conditions, employing linear response theory. From the average dielectric tensor, we computed several macroscopic 
optical properties like the optical absorption coefficient, $\alpha (\omega)$, refractive index, and reflectivity \cite{vaspkit}. 

Figure~\ref{fig4} shows the optical absorption spectra estimated for CAP as a function of incident light wavelength and 
temperature. It is found that $\alpha$ is significantly enhanced under increasing temperature, in some cases by as much 
as an order of magnitude. Similarly to the band gap, the $T$-induced optical absorption variations are largest for 
Ag$_{3}$SeBr (for which $\alpha \sim 10^{3}$--$10^{5}$ at zero temperature and $\sim 10^{4}$--$10^{6}$~cm$^{-1}$ at $200$~K) 
and smallest for Ag$_{3}$SI (for which $\alpha \sim 10^{3}$--$10^{5}$~cm$^{-1}$ at any temperature). It is also noted that 
the most significant optical absorption changes generally occur at low temperatures, that is, within the $0$ to $200$~K 
interval. These $T$-induced $\alpha$ trends align well with the remarkably large influence of EPC on the band gap, 
underscoring the critical role of thermal renormalization effects on the optoelectronic properties of CAP. 

Unfortunately, we cannot directly compare our theoretical $\alpha (\omega)$ results with experimental data, as such data 
is not available in the literature. Notably, Caño \emph{et al.} \cite{cano2024} measured the optical absorption coefficient 
of CAP films scaled by their layer thickness, $d$, specifically, $\overline{\alpha} \equiv \alpha \cdot d$. However, since 
the thickness of the synthesized CAP films was not determined in work \cite{cano2024}, we cannot access the physical quantity 
of interest. In this regard, performing new optoelectronic experiments on CAP films accross a broad range of temperatures, 
including the low-$T$ regime, would be highly desirable. 
\\

{\bf EPC mechanisms in CAP.}~We have already shown that EPC effects are essential for understanding the thermal evolution 
of the optoelectronic properties of CAP and for achieving a consistent agreement between first-principles calculations and 
room-temperature experiments. Next, we focus on unraveling the primary ionic-electronic mechanisms underlying these key 
EPC effects. 

As shown in Fig.~\ref{fig5}a, the influence of each of the fifteen $\Gamma$ phonon modes on the band gap of Ag$_{3}$SBr 
was analysed by monitoring the change in $E_{g}$ driven by frozen-phonon eigenmode distortions of increasing amplitude, 
$u$. The $\Gamma$ phonons were classified into acoustic (A), optical polar (P) and optical nonpolar (NP), where the P 
phonons break the inversion symmetry of the centrosymmetric cubic $Pm\overline{3}m$ phase. It was found that $E_{g}$ is 
unresponsive to acoustic phonon distortions, as expected, while optical P phonons produce the largest band-gap variations. 
As the amplitude of the optical phonon distortions increases, $E_{g}$ systematically decreases in both the P and NP 
cases.

Figure~\ref{fig5}b shows the value of the derivative of the band gap with respect to the phonon distortion amplitude, $u$, 
expressed as a function of the phonon eigenmode energy (as obtained from $T$-renormalised phonon calculations --Methods--). 
We found that low-energy polar phonon modes ($\sim 10$~meV) cause the most significant band-gap reductions, followed by 
high-energy lattice vibrations of the same type ($\sim 200$~meV). At room temperature, phonon excitations with the lowest 
energy host the highest populations and, consequently, represent the most characteristic lattice vibrations in the crystal. 
Therefore, based on the results shown in Figs.~\ref{fig3} and \ref{fig5}, we conclude that low-energy polar phonon modes 
are primarily responsible for the substantial temperature-induced $E_{g}$ reduction reported in this study for CAP compounds. 

The eigenmode of the optical P phonon with the lowest energy is represented in Fig.~\ref{fig5}b. As observed therein, 
this frozen-phonon lattice distortion reduces the distance between the central sulfur atom and one adjacent silver 
atom (Ag2), while increasing the other two S--Ag1 and S--Ag3 bond lengths, compared to the undistorted cubic unit cell. 
Figure~\ref{fig5}c summarizes the relative bond length variation, in absolute value, for all pairs of atoms resulting 
from each of the fifteen $\Gamma$ phonon modes calculated for the cubic $Pm\overline{3}m$ phase. As shown therein, the 
optical P phonons produce the largest S--Ag distance changes (up to $20$\%), while the optical NP phonons cause the 
largest Br--Ag bond length variations (up to $12$\%). The Br--Ag  bond lengths are also appreciably impacted by the 
optical P phonons ($5$--$10$\%). This general behaviour is reminiscent of that observed for optical polar phonons in 
model perovskite oxides like BaTiO$_{3}$ (with atomic substitutions Ag$\leftrightarrow$O, S$\leftrightarrow$Ti and 
Br$\leftrightarrow$Ba) \cite{cohen92,batio3}. 

After identifying the phonon modes that underpin the giant $T$-induced band-gap reduction reported in this study for 
CAP, specifically low-energy optical P modes, we further analyse the induced changes in the electronic band structure. 
Figure~\ref{fig6}a shows the electronic density of states calculated for the archetypal compound Ag$_{3}$SBr 
(equilibrium geometry). It is observed that the top of the valence band (VB) is dominated by highly hybridized silver 
$d$ and chalcohalide $p$ electronic orbitals, while the bottom of the conduction band (CB) is dominated by isotropic 
and more delocalized S and Ag $s$ orbitals. The electronic band structure in Fig.~\ref{fig6}b shows that the VB 
corresponds to the high-symmetry reciprocal space point M~$(1/2,1/2,0)$, while the CB to the center of the Brillouin 
zone, $\Gamma$~$(0,0,0)$, thus the band gap of Ag$_{3}$SBr is indirect (we have checked that the same conclusion applies 
to the rest of CAP analyzed in this study).

The effects on the electronic band structure resulting from a frozen-phonon lattice distortion corresponding to the 
lowest-energy optical P eigenmode are twofold (Fig.~\ref{fig6}b). First, due to the phonon-induced inversion symmetry 
breaking, the energy band degeneracy at the reciprocal space point M is lifted. However, the band gap of the system 
is practically unaffected by this energy degeneracy lifting effect. Second, the CB edge experiences a significant 
decrease and, as a consequence, the band gap of the system is reduced by approximately $30$\%. Therefore, we may conclude 
that the giant $T$-induced $E_{g}$ reduction reported in this study for CAP is primarily caused by low-energy polar 
phonon modes that induce a pronounced CB energy decrease. 

To better understand the electronic origins of the optical P phonon-induced CB energy lowering, we constructed a 
tight-binding (TB) model based on Wannier functions that accurately reproduces our DFT band structure results (Methods 
and Supplementary Fig.1). Specifically, the TB model consists of $s$, $p$, and $d$ orbitals for the five unit-cell
atoms, resulting in a total of $45$ distinct Wannier orbitals. Consistently, the TB model reproduces the dominant 
Ag and S $s$ character of the CB and its energy lowering under the polar lattice distortion of interest (Figs.~\ref{fig6}c--d). 

According to this TB model, the impact of the frozen-phonon distortion on the Ag and S $s$ conduction orbitals is twofold. 
First, the difference in their kinetic energies, corresponding to the diagonal Hamiltonian matrix elements difference 
$|\left< {\rm Ag}~s \right| H \left| {\rm Ag}~s \right> - \left< {\rm S }~s \right| H \left| {\rm S}~s \right>|$, decreases 
(Supplementary Fig.1). And second, the hopping $s$ term involving the Ag2 and S atoms, represented by the off-diagonal 
Hamiltonian matrix element $\left< {\rm Ag2}~s \right| H \left| {\rm S}~s \right>$, increases (Supplementary Fig.1). 
Therefore, the general physical interpretation that follows is that the polar frozen-phonon distortion enhances the 
hybridization of Ag2 and S $s$ conduction orbitals while simultaneously lowering (increasing) the energy level of the 
accompanying bonding (antibonding) state $\sigma$ ($\sigma^{*}$). The revealed EPC mechanism is schematically illustrated 
in Fig.~\ref{fig6}e.

\section*{Discussion}
\label{sec:discussion}
Thermal expansion is another physical mechanism that can influence the band gap of materials \cite{frohlich2}, but it 
was not explicitly considered in this study due to its high computational cost. However, we performed a series of tests 
in which we arbitrarily increased and decreased the unit cell volume of Ag$_{3}$SBr and estimated the corresponding 
relative $E_{g}$ variation (Supplementary Fig.2). We find that, when the volume of the system increases by a reasonable
$\sim 1$\%, as might occur due to thermal expansion near room temperature, the band gap decreases by only $\sim 10$~meV 
(i.e., an order of magnitude smaller change than short-wavelength phonon corrections, Table~\ref{tab:bandgap}). 
Additionally, thermal expansion likely contributes to the softening of optical P modes, increasing their population 
and consequently enhancing the $E_{g}$ reduction due to EPC. Therefore, while thermal expansion effects were omitted 
in our calculations, the reported results likely represent a lower bound of the actual effect, and our conclusions 
for CAP compounds remain robust and accurate.

One may wonder whether, in addition to silver chalcohalide antiperovskites, there may exist other families of materials 
exhibiting similarly large $T$-renormalization effects on the band gap and optical absorption coefficient. As discussed 
in previous sections, the polar nature of low-energy optical phonons appears to be essential in this regard. Consequently, 
a tentative set of necessary conditions for identifying potential materials that display similar $T$-induced effects on 
the optoelectronic properties may include dielectric materials exhibiting (1)~centrosymmetric crystalline phases, 
(2)~low-energy or even imaginary optical polar phonons, and (3)~highly hybridized and delocalized electronic orbitals near 
the Fermi energy level. The availability of large DFT calculations and phonon databases may enable high-throughput materials 
screenings of such a kind \cite{materialsproject,togophonons}. 

Ferroelectric oxide perovskites, exemplified by the archetypal compounds SrTiO$_{3}$ (STO) and BaTiO$_{3}$ (BTO), appear to 
satisfy the set of necessary conditions outlined above. Notably, a significant band-gap modulation has been reported for STO 
under biaxial strain conditions, although this phenomenon arises from different physical mechanisms than those identified 
in this study for CAP, namely, energy degeneracy lifting due to symmetry breaking \cite{sto}. Moreover, the 
experimental room-temperature band gap of BTO ($\approx 3.2$~eV \cite{btoeg}) shows substantial disagreement with 
zero-temperature theoretical estimates obtained with hybrid functionals ($\approx 4.0$~eV \cite{btodfteg}), highlighting 
an experiment-theory inconsistency similar to that described for Ag$_{3}$SBr and Ag$_{3}$SI in the Introduction. Additionally, 
the band gap of the multiferroic oxide perovskite BiFeO$_{3}$ exhibits a remarkable temperature-dependent shrinkage, decreasing 
by approximately $50$\%~ within the temperature range $300 \le T \le 1200$~K \cite{bfo}, likely influenced by the magnetic 
degrees of freedom \cite{bfo2}. These findings suggest that the electron-phonon coupling mechanisms revealed in this study 
for CAP compounds may have broader applicability, extending to other families of renowned functional materials. 

The polar nature of the optical phonon modes, which cause the significant $T$-induced reduction in the $E_{g}$ of CAP, opens 
up exciting technological possibilities. Similar to how an electric field can stabilize a polar phase with ferroelectric 
polarization over a paraelectric phase at constant temperature through a displacive transformation, it is likely that polar 
optical phonons in CAP can also be stimulated using external electric fields. This means that the optoelectronic properties 
of CAP could be efficiently controlled by applying an electric field instead of changing the temperature, offering a more 
practical method for developing advanced optical devices and other technological applications.

Finally, advances in light sources and time-resolved spectroscopy have made it possible to excite specific atomic vibrations 
in solids and to observe the resulting changes in their electronic and electron-phonon coupling properties \cite{opticalc1,
opticalc2,opticalc3}. These developments also suggest the possibility of tuning the optoelectronic properties of CAP, as well 
as of similar materials like oxide perovskites \cite{example1,example2,example3}, through specific phonon excitations using 
optical means such as lasers. This approach may simplify the design and manufacture of practical setups by eliminating the 
need for electrode deposition. Therefore, the results presented in this work are significant not only from a fundamental 
perspective but also for envisioning potential technological applications in which the optical and electronic properties of 
materials could be effectively tuned by external fields and photoexcitation.

\section*{Conclusions}
\label{sec:conclusion}
In this study, we have explored the temperature effects on the band gap of silver chalcohalide antiperovskites (CAP), 
specifically Ag$_{3}$XY compounds (X = S, Se; Y = Br, I), which are promising for energy applications due to their 
lead-free composition, high ionic conductivity, and optoelectronic properties. The key findings of our research are 
summarized as follows. A giant reduction in the band gap of CAP materials has been disclosed at room temperature, 
ranging from 20\% to 60\% relative to their values calculated at zero temperature (neglecting zero-point corrections). 
This large band-gap renormalization brings theoretical predictions closer to experimental results, resolving previous 
discrepancies. 

The significant $E_{g}$ reduction is attributed to strong electron-phonon coupling driven by low-energy polar phonon 
modes, which distort the lattice symmetry, increase the overlap between silver and chalcogen $s$ orbitals in the conduction 
band, and lower the reference energy of the resulting bonding state. With increasing temperature, the optical absorption 
coefficient of CAP materials also rises, enhancing their response to visible light by nearly an order of magnitude and 
highlighting their potential for optoelectronic applications.

This research demonstrates that CAP compounds exhibit giant band-gap renormalization primarily due to electron-phonon 
coupling effects, suggesting that they could be tailored for specific applications through thermal, electric field, and/or 
optical control. These fundamental findings open possibilities for the design of innovative optoelectronic devices and 
establish a foundation for exploring similar effects in other dielectric materials with strong electron-phonon coupling.

\section*{Methods}
\label{sec:methods}
{\bf Zero-temperature first-principles calculations.}~The DFT calculations \cite{vasp,paw,cazorla17} were performed using 
the semilocal PBEsol approximation \cite{pbesol}, considering the following electronic states as valence: Ag $5s$--$4d$, 
S $3s$--$3p$, Se $4s$--$4p$, Br $4s$--$4p$, I $5s$--$5p$. Wave functions were represented in a plane-wave basis set 
truncated at $650$~eV. Using these parameters and a dense {\bf k}-point grid of $8 \times 8 \times 8$ for reciprocal-space 
Brillouin zone (BZ) sampling, we obtained zero-temperature energies converged to within $0.5$~meV per formula unit. For the 
geometry relaxations, a force tolerance of $0.005$~eV$\cdot$\AA$^{-1}$ was imposed on all the atoms. The optoelectronic 
properties were estimated using hybrid functionals \cite{hse06} (Supplementary Fig.3 and Supplementary Discussion) and 
considering spin-orbit coupling (SOC) effects, a computational approach named here as HSEsol+SOC. In this case, to make 
the calculations feasible, the energy cutoff and {\bf k}-point grid were slightly reduced to $550$~eV and $6 \times 6 
\times 6$, respectively. Quantum nuclear effects \cite{cazorla17} were disregarded throughout this work.    

All-electron DFT calculations were also performed with the WIEN2k package \cite{WienBlaha} using the local-density approximation 
(LDA) \cite{Perdew1992} to the exchange correlation energy along with the linearized augmented plane wave method (FP-LAPW) 
\cite{Singh,Blaha1990}. The technical parameters for these calculations were a $10 \times 10 \times 10$ ${\bf k}$-point grid 
and a muffin-tin radius equal to $R_{\rm MT} = 7.0 / K_{\rm max}$, where $K_{\rm max}$ represents the plane-wave cutoff.  
Localized energy-resolved Wannier states \cite{Marzari1997} were then obtained for the tight-binding calculations \cite{Wei2002, 
Wei2006,Jiang2023} considering the relevant Hilbert space in the interval $-10 \le E \le 20$~eV around the Fermi energy.
\\

{\bf Finite-temperature first-principles simulations.}~\emph{Ab initio} molecular dynamics (AIMD) simulations were performed 
in the canonical $(N,V,T)$ ensemble, neglecting thermal expansion effects and employing a large simulation cell containing 
$320$ atoms with periodic boundary conditions applied along the three Cartesian directions. The temperature in the AIMD 
simulations was kept fluctuating around a set-point value by using Nose-Hoover thermostats. Newton's equations of motion 
were integrated using the standard Verlet's algorithm with a time step of $1.5 \cdot 10^{-3}$~ps. $\Gamma$-point sampling 
for reciprocal-space integration was employed in the AIMD simulations, which spanned approximately $100$~ps. These 
calculations were performed with the semilocal PBEsol exchange-correlation functional \cite{pbesol}.
\\

{\bf Harmonic phonon calculations.}~The second-order interatomic force constant matrix of all CAP and resulting harmonic phonon 
spectrum were calculated with the finite-differences method as is implemented in the PhonoPy software \cite{phonopy}. Large 
supercells (i.e., $4 \times 4 \times 4$ for the cubic $Pm\overline{3}m$ supercell containing $320$ atoms) and a dense 
{\bf k}-point grid of $3 \times 3 \times 3$ for BZ sampling were employed for the phonon calculations of targeted structures. 
Several numerical tests were conducted that demonstrated the adequacy of the selected {\bf k}-point grid. These calculations 
were performed with the semilocal PBEsol exchange-correlation functional \cite{pbesol}. 
\\

{\bf Anharmonic phonon calculations.}~The DynaPhoPy code \cite{dynaphopy} was used to calculate the anharmonic lattice dynamics 
(i.e., $T$-renormalized phonons) of CAP in the cubic $Pm\overline{3}m$ phase from \textit{ab initio} molecular dynamics (AIMD) 
simulations. A reduced $2 \times 2 \times 2$ supercell and $4 \times 4 \times 4$ {\bf k}-point grid for BZ sampling were employed 
in the AIMD simulations to maintain high numerical accuracy (Supplementary Discussion and work \cite{benitez24}). 

A normal-mode-decomposition technique was employed in which the atomic velocities $\textbf{v}_{jl}(t)$ ($j$ and $l$ represent 
particle and Cartesian direction indexes) generated during fixed-temperature AIMD simulation runs were expressed like:
\begin{equation}
\textbf{v}_{jl} (t) = \frac{1}{\sqrt{N m_{j}}} \sum_{\textbf{q}s}\textbf{e}_{j}(\textbf{q},s) 
	e^{i \textbf{q} \textbf{R}_{jl}^{0}} v_{\textbf{q}s}(t)~,
\label{eq1}
\end{equation}
where $N$ is the number of particles, $m_{j}$ the mass of particle $j$, $\textbf{e}_{j}(\textbf{q},s)$ a phonon mode eigenvector 
($\textbf{q}$ and $s$ stand for the wave vector and phonon branch), $\textbf{R}_{jl}^{0}$ the equilibrium position of particle 
$j$, and $v_{\textbf{q}s}$ the velocity of the corresponding phonon quasiparticle. 

The Fourier transform of the autocorrelation function of $v_{\textbf{q}s}$ was then calculated, yielding the power spectrum:
\begin{equation}
G_{\textbf{q}s} (\omega) = 2 \int_{-\infty}^{\infty} \langle v_{\textbf{q}s}^{*}(0) v_{\textbf{q}s}(t) \rangle e^{i \omega t} dt~. 
\label{eq2}
\end{equation}
Finally, this power spectrum was approximated by a Lorentzian function of the form:
\begin{equation}
G_{\textbf{q}s} (\omega) \approx \frac{\langle |v_{\textbf{q}s}|^{2} \rangle}{\frac{1}{2} \gamma_{\textbf{q}s} 
        \pi \left[ 1 + \left( \frac{\omega - \omega_{\textbf{q}s}}{\frac{1}{2}\gamma_{\textbf{q}s}} \right)^{2}  \right]}~, 
\label{eq3}
\end{equation}
from which a $T$-renormalized quasiparticle phonon frequency, $\omega_{\textbf{q}s} (T)$, was determined as the peak position, 
and the corresponding phonon linewidth, $\gamma_{\textbf{q}s} (T)$, as the full width at half maximum. These calculations were 
performed with the semilocal PBEsol exchange-correlation functional \cite{pbesol}.
\\

{\bf Short-wavelength phonon band-gap correction.}~ The electron-phonon correction to the band gap due to the short-range 
phonon modes was computed as the difference between the band gap at zero temperature for the static structure and the average 
band gap obtained from AIMD simulations performed with a supercell, namely:
\begin{equation}
    \Delta E_{g}^{S} (T) = \lim_{t_{0} \to \infty}\frac{1}{t_{0}}\int_{0}^{t_{0}}E_{g}^{\mathbf{R}(t)}dt - E_{g}(0),
\label{eq:corr-s-b}
\end{equation}
where $\mathbf{R}$ represents the positions of the atoms in the supercell at a given time $t$ of the AIMD simulation. 
This expression can be numerically approximated as shown in Eq.~(\ref{eq:corr-s}), where the band gap is averaged over 
a finite number, $N$, of configurations. Similarly, thermal effects on the dielectric tensor were computed.

These calculations were performed with the hybrid HSEsol exchange-correlation functional \cite{hse06} and considering
spin-orbit coupling effects (HSEsol+SOC). Due to involved high computational expense, the AIMD calculations were performed 
with a supercell containing $40$ atoms and a ${\bf k}$-point mesh of a single point. The total number of configurations 
used for the average was $N=10$ for each material and temperature. These values were found to be appropriate for obtaining 
band-gap results accurate to within $0.1$~eV (Supplementary Discussion).
\\

{\bf Long-wavelength phonon band-gap correction.}~The electron-phonon correction to the band gap due to long-range phonon 
modes was computed using the Fr\"ohlich equation for a three-dimensional polar material \cite{frohlich1,frohlich2,frohlich3,
frohlich4}. This correction was determined as the difference in the shifts of the conduction and valence bands, as given in 
Eq.~(\ref{eq:corr-l}), where the shift of each band was computed using the expression in Eq.~(\ref{eq:corr-l-2}). The physical 
quantities entering this latter expression were determined with DFT methods; the electron and hole effective masses were 
computed using the parabolic approximation at the maximum of the valence band and the minimum of the conduction band, 
respectively. For the LO phonon frequency, we used an effective value computed as the average of the three corresponding
modes, as done in work \cite{melo23}. Temperature-induced anharmonic effects were fully taken into account for the
calculation of these LO phonon frequencies. 

The high-frequency and static dielectric constants, whose values were unusually large in static calculations, were recomputed 
as functions of temperature (using the same approach as we applied to the band gap and dielectric tensor) to capture thermal 
effects. We also considered the anisotropy of the dielectric tensor using the formula:
\begin{equation}
    \frac{1}{\varepsilon_{i}} = Tr(\bm{\varepsilon}_{i}^{-1})/3,
\end{equation}
where $\varepsilon$ represents the dielectric constant associated with the dielectric tensor $\bm{\varepsilon}$.
\\

\section*{Acknowledgements}
We acknowledge financial support by the Spanish Ministry of Science under the Grants No. TED2021-130265B-C22,
TED2021-130265B-C21, No. PID2020-112975 GB-I00, No. RYC2018-024947-I and by the Generalitat de Catalunya under 
the Grants No. 2021SGR-00343 and No. 2021SGR-01519. Computational support was provided by the Red Española de 
Supercomputación under the Grants No. FI-2023-2-0004, FI-2023-3-0043, FI-2024-1-0005 and FI-2024-2-0003. This 
work is part of the Maria de Maeztu Units of Excellence Programme CEX2023-001300-M funded by MCIN/AEI 
(10.13039/501100011033). P.B. acknowledges support from the predoctoral program AGAUR-FI ajuts (2024 FI-1 00070) 
Joan Oró, which is backed by the Secretariat of Universities and Research of the Department of Research and 
Universities of the Generalitat of Catalonia, as well as the European Social Plus Fund. C.L. acknowledges support 
from the Spanish Ministry of Science, Innovation and Universities under an FPU grant. E.S. is grateful to the 
ICREA Academia program. Work at LIST was supported by the Luxembourg National Research Fund through grant 
C21/MS/15799044/FERRODYNAMICS. This project received funding from the European Union's H2020 European Research 
Council under the grant agreement number 866018 (SENSATE). S.C. and B.M. are supported by a EPSRC grant [EP/V062654/1]
and R.J. and B.M. are supported by a UKRI Future Leaders Fellowship [MR/V023926/1]. B.M. also acknowledges support 
from the Gianna Angelopoulos Programme for Science, Technology, and Innovation, and from the Winton Programme for 
the Physics of Sustainability.


\begin{thebibliography}{30}

\bibitem{epc1} Giustino, F.
	       Electron-phonon interactions from first principles.
		\textit{Rev. Mod. Phys.} \textbf{89}, 015003 (2017).

\bibitem{epc2} Lin, Z., Zhigilei, L.V. and Celli, V.
	       Electron-phonon coupling and electron heat capacity of metals under conditions of strong electron-phonon 
	       nonequilibrium.
	       \textit{Phys. Rev. B} \textbf{77}, 075133 (2008).	

\bibitem{epc3} Bohnen, K.-P., Heid, R. and Renker, B.
	       Phonon dispersion and electron-phonon coupling in MgB$_{2}$ and AlB$_{2}$.
	       \textit{Phys. Rev. Lett.} \textbf{86}, 5771 (2001).	

\bibitem{monserrat18} Monserrat, B., Park, J.-S. and Walsh, A.
	              Role of electron-phonon coupling and thermal expansion on band gaps, carrier mobility, and 
		      interfacial offsets in kesterite thin-film solar cells.
	              \textit{Appl. Phys. Lett.} \textbf{112}, 193903 (2018).

\bibitem{monserrat17} Monserrat, B. and Vanderbilt, D.
	              Temperature dependence of the bulk Rashba splitting in the bismuth tellurohalides.
		      \textit{Phys. Rev. Mater.} \textbf{1}, 054201 (2017).

\bibitem{varshni67} Varshni, Y. P.
	            Temperature dependence of the energy gap in semiconductors.
		    \textit{Physica} \textbf{34}, 149 (1967).

\bibitem{ahc1} O'Donnell, K. P. and Chen, X.
	       Temperature dependence of semiconductor band gaps.
	       \textit{Appl. Phys. Lett.} \textbf{58}, 2924 (1991).	

\bibitem{ahc2} Ponc\'e, S., Gillet, Y., Janssen, J.L., Marini, A., Verstraete, M. and Gonze, X.
	       Temperature dependence of the electronic structure of semiconductors and insulators.
	       \textit{J. Chem. Phys.} \textbf{143}, 102813 (2015).	

\bibitem{ahc3} Park, J., Saidi, W.A., Chorpening, B. and Duan, Y.
	       Applicability of Allen-Heine-Cardona theory on MO$_{x}$ metal oxides and ABO$_{3}$ perovskites: 
	       Toward high-temperature optoelectronic applications.
               \textit{Chem. Mater.} \textbf{34}, 6108 (2022).

\bibitem{diamond1} Giustino, F., Louie, S.G. and Cohen, M. L. 
	           Electron-phonon renormalization of the direct band gap of diamond. 
		   \textit{Phys. Rev. Lett.} \textbf{105}, 265501 (2010).

\bibitem{diamond2} Monserrat, B., Drummond, N.D. and Needs, R.J.
	           Anharmonic vibrational properties in periodic systems: energy, electron-phonon coupling, and stress. 
		   \textit{Phys. Rev. B} \textbf{87}, 144302 (2013).

\bibitem{antimony} Liu, Y., Monserrat, B. and Wiktor, J. 
	           Strong electron-phonon coupling and bipolarons in Sb$_{2}$S$_{3}$. 
		   \textit{Phys. Rev. Mater.} \textbf{7}, 085401 (2023).

\bibitem{molecular} Monserrat, B., Engel, E.A. and Needs, R.J. 
	            Giant electron-phonon interactions in molecular crystals and the importance of nonquadratic coupling.
                    \textit{Phys. Rev. B} \textbf{92}, 140302 (2015).

\bibitem{anomal1} Villegas, C.E.P., Rocha, A.R. and Marini, A.
	          Anomalous temperature dependence of the band gap in black phosphorus.
		  \textit{Nano Lett.} \textbf{16}, 5095 (2016).

\bibitem{anomal2} Saidi, W.A., Ponc\'e, S. and Monserrat, B.
	          Temperature dependence of the energy levels of methylammonium lead iodide perovskite from first-principles.
		  \textit{J. Phys. Chem. Lett.} \textbf{7}, 5247 (2016).

\bibitem{anomal3} Artus, L. and Bertrand, Y.
	          Anomalous temperature dependence of fundamental gap of AgGaS$_{2}$ and AgGaSe$_{2}$ chalcopyrite compounds.
		  \textit{Sol. Stat. Commun.} \textbf{61}, 733 (1987).

\bibitem{benitez24} Ben\'itez, P., L\'opez, C., Liu, C., Caño, I., Tamarit, J.-Ll., Saucedo, E. and Cazorla, C.
                    Crystal structure prediction and phase stability in highly anharmonic silver-based chalcohalide 
		    anti-perovskites.
		    arXiv:2406.04966 (2024).

\bibitem{takahashi66} Takahashi, T. and Yamamoto, O.
                      The Ag/Ag$_{3}$SI/I$_{2}$ solid-electrolyte cell.
                      \textit{Electrochim. Acta} \textbf{11}, 779 (1966).

\bibitem{hull04} Hull, S.
                 Superionics: crystal structures and conduction processes.
                 \textit{Rep. Prog. Phys.} \textbf{67}, 1233 (2004).

\bibitem{wakamura90} Wakamura, K., Miura, F., Kojima, A. and Kanashiro, T.
                     Observation of anomalously increasing phonon damping constant in the $\beta$ phase of the fast-ionic
                     conductor Ag$_{3}$SI.
                     \textit{Phys. Rev. B} \textbf{41}, 2758 (1990).

\bibitem{sakuma85} Sakuma, T.
                   Treatment of anharmonic thermal vibration by using transformation of scattering vector.
                   \textit{J. Phys. Soc. Jpn.} \textbf{54}, 4188 (1985).

\bibitem{kawamura80} Kawamura, J., Shimoji, M. and Hoshino, H.
                     The ionic conductivity and thermoelectric power of the superionic conductor Ag$_{3}$SBr.
                     \textit{J. Phys. Soc. Jpn.} \textbf{50}, 194 (1981).

\bibitem{magistris72} Magistris, A., Pezzati, E. and Sinistri, C.
                      Thermoelectric properties of high-conductivity solid electrolytes.
                      \textit{Z. Naturforsch.} \textbf{27a}, 1379 (1972).

\bibitem{palazon22} Palazon, F.
                    Metal chalcohalides: Next generation photovoltaic materials?
                    \textit{Sol. RRL} \textbf{6}, 2100829 (2022).

\bibitem{ghorpade23} Ghorpade, U.V., Suryawanshi, M. P., Green, M. A., Wu, T., Hao, X. and Ryan, K. M.
                     Emerging chalcohalide materials for energy applications.
                     \textit{Chem. Rev.} \textbf{123}, 327 (2023).

\bibitem{luna23} Sebastiá-Luna, L., Rodkey, N., Mirza, A. S., Mertens, S., Lal, Melchor, A., Carranza, G.,
                 Calbo, J., Righetto, M., Sessolo, M., Herz, L. M., Vandewal, K., Ortí, E., Morales-Masis, M.,
                 Bolink, H. J. and Palazon, F.
                 Chalcohalide antiperovskite thin films with visible light absorption and high charge-carrier mobility
                 processed by solvent-free and low-temperature methods.
                 \textit{Chem. Mater.} \textbf{35}, 6482 (2023).

\bibitem{cano2024} Caño, I., Turnley, J. W., Benítez, P., López-Álvarez, C., Asensi, J.-M., Payno, D., Puigdollers, J.,
                 Placidi, M., Cazorla, C., Agrawal, R. and Saucedo, E.
                 Novel synthesis of semiconductor chalcohalide anti-perovskites by low-temperature molecular precursor
                 ink deposition methodologies.
                 \textit{J. Mater. Chem. C} \textbf{12}, 3154 (2024).

\bibitem{liu21} Liu, Z., Mi, R., Ji, G., Liu, Y., Fu, P., Hu, S., Xia, B. and Xiao, Z.
		Bandgap engineering and thermodynamic stability of oxyhalide and chalcohalide antiperovskites. 
		\textit{Ceram. Int.} \textbf{47}, 32634 (2021).

\bibitem{hse06} Krukau, A. V., Vydrov, O. A., Izmaylov, A. F., Scuseria, G. E.
                Influence of the exchange screening parameter on the performance of screened hybrid functionals.
                \textit{J. Chem. Phys.} \textbf{125}, 224106 (2006).

\bibitem{sakuma80} Sakuma, T. and Hoshino, S.
                   The phase transition and the structures of superionic conductor Ag$_{3}$SBr.
                   \textit{J. Phys. Soc. Jpn.} \textbf{49}, 678 (1980).

\bibitem{hoshino81} Hoshino, S., Fujishita, H., Takashige, M. and Sakuma, T.
                    Phase transition of Ag$_{3}$SX (X= I, Br).
                    \textit{Solid State Ion.} \textbf{3}, 35 (1981).

\bibitem{cho94} Cho, N., Kikkawa, S., Kanamaru, F. and Yoshiasa, A.
                Structural refinement of Ag$_{3}$SI by single crystal X-ray diffraction method.
                \textit{Solid State Ion.} \textbf{68}, 57 (1994).

\bibitem{frohlich1} Zacharias, M., Scheffler, M. and Carbogno, C.. 
	            Fully anharmonic nonperturbative theory of vibronically renormalized electronic band structures. 
		    \textit{Phys. Rev. B} \textbf{102}, 045126 (2020).

\bibitem{frohlich2} Chen, S., Parker, I. J. and Monserrat, B.
		    Temperature effects in topological insulators of transition metal dichalcogenide monolayers. 
		    \textit{Phys. Rev. B} \textbf{109}, 155125 (2024).

\bibitem{frohlich3} Poncé, S., Gillet, Y., Janssen, J. L., Marini, A., Verstraete, M. and Gonze, X. 
	            Temperature dependence of the electronic structure of semiconductors and insulators.
                    \textit{J. Chem. Phys.} \textbf{143}, 102813 (2015).

\bibitem{frohlich4} Zacharias, M. and Giustino, F. 
	            Theory of the special displacement method for electronic structure calculations at finite temperature. 
		    \textit{Phys. Rev. Res.} \textbf{2}, 013357 (2020). 

\bibitem{melo23} Melo, P. M., Abreu, J. C., Guster, B., Giantomassi, M., Zanolli, Z., Gonze, X. and Verstraete, M. J. 
	         High-throughput analysis of Fr\"ohlich-type polaron models. 
		 \textit{npj Comput. Mater.} \textbf{9}, 147 (2023).

\bibitem{vaspkit} Wang, V., Xu, N., Liu, J. C., Tang, G., Geng, W. T. 
	          VASPKIT: A user-friendly interface facilitating high-throughput computing and analysis using VASP code. 
		  \textit{Comput. Phys. Commun.} \textbf{267}, 108033 (2021).

\bibitem{cohen92} Cohen, R. E.
	          Origin of ferroelectricity in perovskite oxides.
		  \textit{Nature} \textbf{358}, 136 (1992).

\bibitem{batio3} Zhong, W., Vanderbilt, D. and Rabe, K. M.
	         Phase transitions in BaTiO$_{3}$ from first principles.
		 \textit{Phys. Rev. B} \textbf{73}, 1861 (1994).

\bibitem{materialsproject} Jain, A., Ong, S. P., Hautier, G. \emph{et al.}
                           The Materials Project: A materials genome approach to accelerating materials 
			   innovation.
			   \textit{APL Mater.} \textbf{1}, 011002 (2013).

\bibitem{togophonons} https://github.com/atztogo/phonondb 

\bibitem{sto} Berger, R. F., Fennier, C. J. and Neaton, J. B.
              Band gap and edge engineering via ferroic distortion and anisotropic strain: The case of SrTiO$_{3}$.
              \textit{Phys. Rev. Lett.} \textbf{107}, 146804 (2011).

\bibitem{btoeg} S.H. Wemple. 
 	        Polarization fluctuations and the optical-absorption edge in BaTiO$_{3}$.
                \textit{Phys. Rev. B} \textbf{2}, 2679 (1970). 

\bibitem{btodfteg} Evarestov, R. A. and Bandura, A. V.
	           First-principles calculations on the four phases of BaTiO$_{3}$.
                   \textit{J. Comput. Chem.} \textbf{33}, 1123 (2012).

\bibitem{bfo} Weber, M. C., Guennou, M., Toulouse, C., Cazayous, M., Gillet, Y., Gonze, X. Kreisel, J.
	      Temperature evolution of the band gap in BiFeO$_{3}$ traced by resonant Raman scattering.
	      \textit{Phys. Rev. B} \textbf{93}, 125204 (2016).

\bibitem{bfo2} Cazorla, C. and Íñiguez-González, J.
	       Insights into the phase diagram of bismuth ferrite from quasiharmonic free-energy calculations.
	       \textit{Phys. Rev. B} \textbf{88}, 214430 (2013).

\bibitem{opticalc1} Kennes, D., Wilner, E., Reichman, D. \emph{et al.} 
		    Transient superconductivity from electronic squeezing of optically pumped phonons. 
		    \textit{Nat. Phys.}  \textbf{13}, 479 (2017).

\bibitem{opticalc2} Dekorsky, T., K\"utt, W., Pfeifer, T. and Kurz, H.
	            Coherent control of LO-phonon dynamics in opaque semiconductors by femtosecond laser pulses.
		    \textit{Europhys. Lett.} \textbf{23}, 223 (1993).

\bibitem{opticalc3} Pomarico, E., Mitrano, M., Bromberger, H. \emph{et al.}
	            Enhanced electron-phonon coupling in graphene with periodically distorted lattice.
		    \textit{Phys. Rev. B}  \textbf{95}, 024304 (2017).

\bibitem{example1} Qi, Y., Liu, S., Lindenberg, A. M. and Rappe, A. M.
	           Ultrafast electric field pulse control of giant temperature change in ferroelectrics.
		   \textit{Phys. Rev. Lett.} \textbf{120}, 055901 (2018).

\bibitem{example2} Peng, B., Hu, Y., Murakami, S., Zhang, T. and Monserrat, B. 
	           Topological phonons in oxide perovskites controlled by light.
		   \textit{Sci. Adv.} \textbf{6}, eabd1618 (2020).

\bibitem{example3} Rurali, R., Escorihuela-Sayalero, C., Tamari, J.-Ll., Íñiguez-González, J. and Cazorla, C.
	           Giant photocaloric effects across a vast temperature range in ferroelectric perovskites.
		   \textit{Phys. Rev. Lett.} \textbf{133}, 116401 (2024).

\bibitem{vasp} Kresse, G. and Furthm\"uller, J. 
               Efficient iterative schemes for ab initio total-energy calculations using a plane-wave basis set.
	       \textit{Phys. Rev. B} \textbf{54}, 11169 (1996).

\bibitem{paw} Bl\"ochl, P. E. 
              Projector augmented-wave method.
	      \textit{Phys. Rev. B} \textbf{50}, 17953 (1994).

\bibitem{cazorla17} Cazorla, C. and Boronat, J.
                    Simulation and understanding of atomic and molecular quantum crystals.
                    \textit{Rev. Mod. Phys.} \textbf{89}, 035003 (2017).

\bibitem{pbesol} Perdew, J. P., Ruzsinszky, A., Csonka, G. I. \textit{et al.}
	         Restoring the density-gradient expansion for exchange in solids and surfaces.
	         \textit{Phys. Rev. Lett.} \textbf{100}, 136406 (2008). 

\bibitem{WienBlaha} Blaha, P., Karlheinz, S., Fabien, T., Laskowski, R., Georg, M. and Laurence, D. M. 
	            WIEN2k: An APW+lo program for calculating the properties of solids.
                    \textit{The Journal of Chemical Physics} \textbf{152}, 074101 (2020).

\bibitem{Perdew1992} Perdew, J. P. and Wang, Y.
		     Accurate and simple analytic representation of the electron-gas correlation energy.
                     \textit{Phys. Rev. B} \textbf{45}, 13244 (1992).

\bibitem{Singh} David, J. S.   
                Planewaves, pseudopotentials and the LAPW method.
                \textit{Springer New York, NY} (2006).

\bibitem{Blaha1990} Blaha, P., Schwarz, K., Sorantin, P. and Trickey, S. B.
	            Full-potential, linearized augmented plane wave programs for crystalline systems.
                    \textit{Comput. Phys. Commun.} \textbf{59}, 399 (1990).

\bibitem{Marzari1997} Marzari, N. and Vanderbilt, D.
		      Maximally localized generalized Wannier functions for composite energy bands.
                      \textit{Phys. Rev. B} \textbf{56}, 12847 (1997).

\bibitem{Wei2002} Wei, K., Rosner, H., Pickett, W. E. and Scalettar, R. T.
		  Insulating ferromagnetism in La$_{4}$Ba$_{2}$Cu$_{2}$O$_{10}$: An ab initio Wannier function analysis.
                    \textit{Phys. Rev. Lett.} \textbf{89}, 167204 (2002).

\bibitem{Wei2006} Wei-Guo, Y., Volja, D. and Wei, K. .
		  Orbital ordering in LaMnO${_3}$: Electron-electron versus electron-lattice interactions.
                    \textit{Phys. Rev. Lett.} \textbf{96}, 116405 (2006).

\bibitem{Jiang2023} Ruoshi, J., Lang, Z.-J., Berlijn, T. and Ku, W.
		    Variation of carrier density in semimetals via short-range correlation: A case study with nickelate 
		    NdNiO$_{2}$.
                    \textit{Phys. Rev. B} \textbf{108}, 155126 (2023).

\bibitem{phonopy} Togo, A. and Tanaka, I.  
                  First principles phonon calculations in materials science.
                  \textit{Scr. Mater} \textbf{108}, 1 (2015).

\bibitem{dynaphopy} Carreras, A., Togo, A. and Tanaka, I.
                    DynaPhoPy: A code for extracting phonon quasiparticles from molecular dynamics simulations.
                    \textit{Comput. Phys. Commun.} \textbf{221}, 221 (2017).


\end{thebibliography}
\end{document}